\documentstyle[12pt]{article}
\begin{document}
\baselineskip=24pt
\newcommand{\di}{\displaystyle}
\noindent{\Large \bf Unique Deformation of Local Quantum Field Theory
Resulting in Divergence-free \\
Amplitudes}

\vspace{1cm}
\baselineskip=15pt
\noindent{\bf G.H. GADIYAR}\\
\noindent{\it Theory Division, Bharatiya Vigyan Sangh,\\
Plot 108, First Main Road, Ram Nagar North Extension,\\
Velachery, Madras 600 042 INDIA.\\
e-mail: padma@imsc.ernet.in}

\vspace{2cm}
\baselineskip=24pt
\noindent{\bf Abstract.} An essentially unique deformation of the product
of quantum fields at the same spacetime point is obtained. It is proposed to
replace local quantum field theory with another structure which uses
a $*$-product. The resulting theory contains a fundamental length and
is free from divergences. This provides the third deformation suggested
by Faddeev.


 
\newpage
Quantum Field Theory (QFT) is plagued with the problem of divergences.
The problem is with the product of local fields at the same spacetime point.
In Bogoliubov's program [2] the ambiguity in the time-ordered product at
coincident spacetime point is made use of to get finite S-matrix elements
consistent with Lorentz covariance, locality, causality and unitarity.
In the class of so called renormalizable theories the finite S-matrix 
elements depend only on a finite number of independent parameters.
Unfortunately Einstein gravity does not belong to this class. There is
a widely held belief that the problems of divergences in local QFT and
quantized gravity are inter-related. A correct quantization of gravity
may automatically remove divergences of QFT. This is all the more probable
because Newton's gravitational constant provides for fundamental length
scale that is the Planck length.

Faddeev [3] (see also [4]) has observed that the special theory of relativity and quantum
mechanics which replaced Galilean relativity at classical mechanics can
be interpreted as deformations [1], [5], [6] of the earlier structures. In the process
two fundamental parameters $c$ the velocity of light and $\hbar$ the
Planck constant enter physics. $c$ provides a cutoff for velocities and
$\hbar$ for phase-space. Faddeev has proposed that another deformation
bringing in a fundamental length is perhaps required. This will provide
a system of fundamental or natural units for all dimensionful parameters
involving mass, length and time.

In this article it is shown that there is an essentially unique deformation of the product of fields at the same spacetime point preserving associativity. 
Surprisingly it turns out that the deformation parameter provides an 
in-built momentum cutoff. The proposal here provides a natural deformation
of local QFT thus implementing  Faddeev's program. 

There have been many attempts to quantize spacetime and include a 
fundamental length by making coordinates non-commutative [5]. However
this procedure is not unique and has many ad hoc assumptions and has
many difficulties. In contrast the approach here is to consider 
deformation of the ring of functions on spacetime. This is the dual object
in the mathematical sense. It is well known that deformation is natural
in this dual object. For example, in the case of quantum mechanics
the deformation of Poisson bracket to Moyal bracket [1], [8]  works with functions
on phase-space.

As mentioned earlier the problem of divergences in QFT is related to the
product  of local fields at the same spacetime point. Consider now a 
possible deformation preserving the associativity of such products.
As is conventional the deformed product is denoted by $*$. For functions
of one variable the solution is
\begin{eqnarray*}
f(x) * g(x) &=& \sum_{n=0}^{\infty} \frac{L^{2n}}{n!} \, \frac{ d^n f}{dx^n} \, \frac{d^ng}{dx^{n}}\\
&=& f(x) \, e^{{\di L^2 \frac{d}{\overleftarrow{dx}} \frac{d}{{\overrightarrow{dx}}}}} \, g(x) \, . \hspace{6.5cm} (1)
\end{eqnarray*}
In order to prove the uniqueness and associativity consider first
$$
e^{ikx}*e^{ik'x} = f(k,k') e^{i(k+k')x} \, . \eqno(2)
$$
The requirement of associativity is
$$
e^{ikx}*(e^{ik'x}* e^{ik''x}) ~=~ (e^{ikx}*e^{ik'x})* e^{ik''x} \, . \eqno(3) 
$$
This implies that
$$
f(k, k'+k") f(k', k'') ~=~ f(k, k')f(k+k', k'') \, . \eqno(4)
$$
This functional equation has unique nontrivial solution 
$$
f(k, k') ~=~ e^{L^2 kk'} \eqno(5)
$$
where $L^2$ is some real or complex constant. By Fourier transforming $f(x)$ 
and $g(x)$ and using equation (5), equation (1) follows.

Now the results are applied to local QFT. For the present consider
Euclidean fields. It is well known that once the Euclidean field theory
is well defined the corresponding Minkowski theory can be got uniquely
by analytic continuation. The product of local fields is now replaced by
the $*$-product which has a straight-forward generalization to four dimensions.
$$
f(x) * g(x) = f(x) e^{{\di L^2 \eta_{\mu \nu}\frac{\partial}{\partial {\overleftarrow x}_{\mu}} \frac{\partial}{\partial {\overrightarrow x}_{\nu}}}} g(x) \, . \eqno(6)
$$
Using the Fourier transforms, 
$$
f(x) * g(x) = \int \frac{d^4 k}{(2\pi)^4}\, \frac{d^4 k'}{(2\pi)^4} \, \tilde{f}(k) \, \tilde{g}(k') \, e^{L^2 k.k'} \, e^{i(k+k').x} \, . \eqno(7)
$$ 
In particular 
$$
\int d^4x f(x) * g(x) = \int \frac{d^4k}{(2\pi)^4} \, {\tilde f}(k) \, {\tilde g}(-k) \, e^{-L^2 k^2} \, . \eqno(8)
$$
If $L$ is chosen to be a real positive constant it follows 
that high frequency components of the quantum fields are cutoff through
the deformation parameter $L$. This is exactly in analogy with $c$ which
provides a cutoff for velocity in relativity and $\hbar$ which provides
a cutoff in quantum mechanics for phase-space. In this sense this proposal
provides an exact realization of Faddeev's program for obtaining a third
deformation introducing a fundamental length into physical theories. 

It is proposed that all local products of fields be replaced by $*$-products.
Thus the functional integral for a Euclidean scalar field theory is
$$
\int {\cal D} \phi e^{-\frac{1}{2}{\di \int d^4x \partial _{\mu} \phi (x) *
\partial _{\mu} \phi (x) + m^2 \phi (x) * \phi (x) + \lambda \phi (x) *
\phi (x) * \phi (x) * \phi (x)}} \eqno(9)
$$
Now all computations becomes straight-forward going to momentum space in
using formula (8). In particular the free propagator is 
${\di \frac{e^{-k^2 L^2}} {k^2 + m^2}}$. This shows that an automatic cutoff
is provided by $L$. An added advantage of this scheme is that calculations
can be made in the conventional way after the introduction of the $*$-product.

A physical interpretation of the $*$-product is now given. Note the exact
analogy of the $*$-product with Wick's theorem.
$$
: e^{ik\phi} : : e^{ik' \phi} : ~=~ e^{-kk' \Delta} : e^{i(k+k')\phi}:
\eqno(10)
$$
where $\Delta$ is $<\phi \phi>$ which is the covariance for the random
fields $\phi$. Thus the deformation appears to make $x$ a random variable
with covariance $L^2$. In other words, $(\Delta x)^2 \ge L^2$. It is
this graininess of spacetime at length scale $L$ which is providing an
ultra-violet regularization. To illustrate this relationship between 
ordinary product and $*$-product note that 
$$
x_{\mu} * x_{\nu} = x_{\mu} x_{\nu} + \eta_{\mu \nu} L^2 \, . \eqno(11)
$$
Thus commutativity of the coordinates is not changed in contrast to other
proposals but the light cone is smoothened out. 

The fundamental length $L$ also enters quantum gravity and makes it 
meaningful. The invariance under general coordinate transformation
plays a crucial role in the general theory of relativity. Note that 
this principle has to be drastically modified once ordinary products
are replaced by $*$-products. There is an easy and systematic way of 
guessing what this new principle is to be. In the field theory
approach pioneered by Feynman, Schwinger, Gupta, Thirring, Weinberg and Deser (see discussions in [7]) flat spacetime and Lorentz invariance continue to remain.
It is merely required that there is a symmetric rank two tensor which couples
to the energy-momentum tensor of all fields including itself. 
It is known that this gives the Einstein - Hilbert action of the interaction
of these spin two fields. Thus general coordinate invariance and
Einstein gravity are a consequence of QFT with spin two gravitons
coupling to the energy momentum tensor in flat spacetime. The same
strategy can be adopted in the present approach, however the local
products should be replaced by $*$-products. A unique theory with a 
geometric interpretation will follow. This will be pursued elsewhere.

\vspace{1cm}
\noindent{\bf Acknowledgements} 
\newline Thanks are due to Prof. H.S. Sharatchandra
 for help with the manuscript and the arguments of equation (4) 
and the physical interpretation given in equation (10).

\vspace{1cm}
\noindent{\bf References}
\baselineskip=15pt
\begin{description}
\item{1.} Bayen, F., Flato,M., Fronsdal, C., Lichnerowicz, A. and
Sternheimer, D.: Deformation theory and Quantization I. Deformation
of symplectic structures; II. Physical Applications, {\it Ann.
Phys. NY}. {\bf 111} (1978), 61-110; 111 - 151.
\item{2.} Bogoliubov, N.N. and Shirkov, D.V.: {\it Introduction to the theory
of quantized fields}, Interscience Publishers Inc., New York, 1959.
\item{3.} Faddeev, L.D.: On the relationship between Mathematics and Physics, {\it Asia Pacific Physics News}. {\bf 3} June-July, (1988), 21-22.
\item{4.} Flato, M.: Deformation view of physical theories, {\it 
Czechoslovak J. Phys.} {\bf B32} (1982), 472-475.
\item{5.} Garay, L.J.: Quantum gravity and minimum length, {\it Int.
J. Mod. Phys. A}. {\bf 10} (1995), 145-198.
\item{6.} Gerstenhaber, M.: On the deformation of rings and algebras,
{\it Ann. Math.} {\bf 79} (1964), 59 - 103. 
\item{7.} Isham, C.J., Penrose, R. and Sciama, D.W. (eds): {\it Quantum gravity, an Oxford Symposium}, Oxford University Press, 1975.  
\item{8.} Moyal, J.E.: Quantum mechanics as a statistical theory,
{\it Proc. Camb. Phil. Soc.} {\bf 45} (1949), 99-124.
\end{description}
\end{document}